\documentclass{ptephy_v1}

\preprintnumber{1903.01663} 

\newcommand {\uPIC} {$\mu$-PIC}
\newcommand {\SR} {SR$\mu$-TPC}
\newcommand {{\wire}} {wire $\mu$-TPC}

\begin{document}

\title{Development of a time projection chamber with a sheet-resistor field cage}

\author[1]{Kentaro~Miuchi}
\author[1]{Tomonori~Ikeda}
\author[1]{Hirohisa~Ishiura}
\author[1]{Kiseki~D.~Nakamura}
\author[2]{Atsushi~Takada}
\author[1]{Yasuhiro~Homma}
\author[3,4]{Ko~Abe}
\author[3,4]{Koichi~Ichimura}
\author[1]{Hiroshi~Ito}
\author[3,4]{Kazuyoshi~Kobayashi}
\author[1]{Takuma~Nakamura}
\author[1]{Ryuichi~Ueno}
\author[1]{Takuya~Shimada}
\author[1]{Takashi~Hashimoto}
\author[1]{Ryota~Yakabe}
\author[1]{Atsuhiko~Ochi}

\affil[1]{Department of Physics, Graduate School of Science, Kobe University, 1-1 Rokkodai-cho, Nada-ku, Kobe, Hyogo, 657-8501, Japan \email{miuchi@phys.sci.kobe-u.ac.jp}}
\affil[2]{Division of Physics and Astronomy, Graduate School of Science, Kyoto University, Kitashirakawaoiwake-cho, Sakyo-ku, Kyoto, Kyoto, 606-8502, Japan}
\affil[3]{Kamioka Observatory, Institute for Cosmic Ray Research, the University of Tokyo, Higashi-Mozumi, Kamioka, Hida, Gifu, 506-1205, Japan}
\affil[4]{Kavli Institute for the Physics and Mathematics of the Universe (WPI), the University of Tokyo, 5-1-5 Kashiwanoha, Kashiwa, Chiba, 277-8582, Japan}



\begin{abstract}
A new-concept time projection chamber (TPC) using a commercial resistive sheet, sheet-resistor micro-TPC (\SR),
was developed and its performance was measured. 
{\SR} has the potential
to create a more uniform electric field than conventional TPCs with resistor-chains owing to its continuous sheet resistivity, 
and its production would be easier than that of conventional TPCs.
The material used in this study, Achilles-Vynilas,
was found to be thin, transparent, and low-radioactive.
The performance test with cosmic muons showed 
very promising results, including
the demonstration of a good tracking-performance.
This type of TPC field cage can offer an alternative for the widely used conventional field cages.
\end{abstract}

\subjectindex{H11 Gaseous detectors, H20 Instrumentation for underground experiments}

\maketitle

\section{Introduction}
\label{sec:introduction}

Since the invention of the time projection chamber (TPC) in the 1970's by David Nygren\cite{bib:Nygren}, many types of TPCs have widely been used in various physics experiments to take advantage of their full three-dimensional tracking ability.
Large-scale ($O(\rm1~m)$) TPCs have been developed for 
accelerator experiments (NA49\cite{bib:NA49} and ALICE\cite{bib:ALICE}), and neutrino physics (T2K\cite{bib:T2K}).
Applications of the TPC have been widened to rare-event-search experiments in the 2000's and double-phase liquid noble-gas TPCs, like
LUX\cite{bib:LUX}, PandaX-II\cite{bib:PANDAX}, XENON1T\cite{bib:XENON1T}, and DarkSide-50\cite{bib:DARKSIDE}, are leading the direct dark matter search experiments owing mostly to their particle identification powers. 
A spherical proportional chamber (NEWS-G) with a high-pressure gas has been developed for low-mass dark matter search\cite{bib:NEWS}, as well as a liquid xenon TPC (EXO-200) for the neutrinoless double beta decay search experiments\cite{bib:EXO}.
Furthermore, low-pressure large-volume gaseous TPCs, such as DRIFT-IId\cite{bib:DRIFT}, NEWAGE-0.3b\cite{bib:NEWAGE}, and MIMAC\cite{bib:MIMAC}, have been developed for direction-sensitive dark matter search experiments aiming to detect the tracks of the recoil nuclei. 

Typical field cages for these TPCs consist of field-shaping electrodes (wires, metal plates, aluminized Mylars, etc.) with resistor-chains to supply appropriate potential to the electrodes. The electric fields were precisely calculated
using the finite-element methods so as to achieve a uniform electric field in the detection volume. Each electrode in a typical design was about 1~cm wide with a  0.5~cm spacing in the drift direction. In most cases, the electrodes were designed to have more than 50$\%$ occupancy so as to shield the ground-potential of the vessel, which would cause the deterioration of the electric field. This design, however, causes some non-uniformity of the electric field near the field cage, roughly closer than the width of the electrode. It would be useful if a sheet resistor with continuous resistivity could be used as the electric-field-shaping material because it is expected to form a uniform electric field even in the vicinity of the field cage. Detector assembly is also expected to be relatively easy, and a low background detector can be made without resistors or solders.
Resistive materials with bulk or sheet resistivities have attracted attention
for the development of micro-patterned gaseous detectors (MPGDs)
in terms of discharge suppressions\cite{bib:resistive}\cite{bib:semitron}.
These materials are also widely used for anti-static purposes. In this paper, we report the development and performance measurements
of a sheet-resistor micro-TPC, or the {\SR}.

\section{Detector Assembly}
\label{sec:development}
The material for the field cage of the {\SR} was selected among several candidates.
The required sheet resistivity was $10^{10}-10^{11}~\Omega/\square$ mainly because of the realistic high voltage supply\footnote{A field cage made of a material with a sheet resistivity of $10^{10}~\Omega/\square$ for a 1~m drift length and 1~m diameter has a resistivity of $3\times10^{9}~\Omega$, which requires a current of 10~$\mu$A for a 30~kV voltage.}.
Several commercially-available samples were purchased and their properties were measured. Measured material candidates are summarized in Table~\ref{tab:material}. Semitron, used in Ref.~\cite{bib:semitron} was in the form of a plate, while the others were sheets shipped in rolls.
The sheet resistivities along the widths of the rolls were measured every 10~cm interval\footnote{This was to ensure that there existed at least one direction with a good uniformity, which could be used for the drift direction.}.  
Two copper cubes ($\rm 2\times2\times2~cm^3$) separated by a distance of 2~cm, were used as electrodes for the sheet-resistivity measurement. A bias voltage of up to 1~V was supplied between the electrodes by a dry battery and the current was amplified by a home-made amplifier with a gain of 5~mV/pA. Texas Instruments TLC2652 operational amplifiers were used for the amplifier. 
The bias voltage and current-voltage were read by multimeters (HEWLETT PACKARD 34401A and KEITHLEY 2000 multimeter, respectively) and recorded by a computer. This system was confirmed to be able to measure up to $1.0\times10^{11}~\Omega$ with a precision of 1$\%$ using a standard high precision resistor.

Catalog specifications and the measurement results are listed in Table~\ref{tab:material}.
The error of the sheet resistivity is the standard deviation of the measured resistivities.
For the samples whose measured sheet resistivity varied by more than 
one order of magnitude,
the results are shown with $\sim$ to indicate the orders.
The DPF-arutron showed a resistivity larger than the range of our system, so the lower limit is shown.
In terms of the uniformity of the sheet resistivity, two samples, namely Anti-static PVC (polyvinyl chloride) and Achilles-Vynilas, were found to be relatively good candidates.
It should be noted that
the uniformity does not affect much
for the practical use as an anti-electric material,
thus it is understandable that there was a large variety of the uniformity among the measured products.
Antistatic PVC was found to have a resistivity difference in a large-scale of about 15~$\%$/m while that of Achilles-Vynilas was less than 1$\%$/m.
Achilles-Vynilas was thus selected as the best material among the candidates. One interesting property of this material, which may broaden its application, was that a transmittance of more than 90$\%$ was guaranteed for visible light with a wave length of $>$ 450~nm.

\begin{table}[htbp]
\centering
\caption{\label{tab:material} List of candidates for the field-cage material of the \SR. Catalog specifications and measured sheet resistivities are listed as ''spec.'' and ''meas.'', respectively. The catalog specification of the Semitron ESd was shown in the unit of volume resistivity.}
\smallskip
\begin{tabular}{|l|l|l|l|l|l|}
\hline
product name&material&size&thickness&\multicolumn{2}{|c|}{sheet resistivity}\\ \cline{5-6} 
&&(shipped)&&spec.&meas.\\ \hline 
&&\multicolumn{1}{r|}{[$\rm {m}^2$]}&\multicolumn{1}{r|}{[mm]}&\multicolumn{2}{r|}{[$\times10^{10}~\Omega/\square$]}\\ \hline 
Semitron ESd&	Polyacetal&0.3$\times$0.6&6&$10^8\sim10^{10}$&$1\sim10$ \\
&&&&$\rm(\Omega m)$&(along 0.6~m)\\\hline
Antistatic film&	Polyolefin&1$\times$200&0.05&$<10$&$1\sim100$\\ \hline
Antistatic PVC sheet&PVC&	1.37 $\times$30&0.3&1$\sim$10&$2.0\pm0.2$\\ \hline
Achilles-Vynilas&PVC&1$\times$10&0.2&10&$3.3\pm0.3$\\ \hline
DPF-arutoron&PVC&1.83$\times$50&0.1&51&$>100$\\ \hline
\end{tabular}
\end{table}

The radioactive contamination of the Achilles-Vynilas was measured to ensure its viability for rare-event-search experiments.
A high-purity germanium detector (HPGe)
at the Kamioka observatory in the Kamioka Mine (2,700~m water equivalent) was used. Details of the detector system can be found in Ref.~\cite{bib:HPGe}. The results are summarized in Table~\ref{tab:HPGeresults}.
No finite value was detected and upper limits at 90$\%$ C.L. are shown in Table~\ref{tab:HPGeresults}. Radioactivities of resistors listed in Ref.~\cite{bib:HPGe} are also shown in Table~\ref{tab:HPGeresults} for reference.   
Typical amounts needed for a 10$\times$10$\times$10~${\rm cm^{3}}$-sized TPC
  are $O(\rm 10~g)$ and $O(\rm 0.1\sim1~g)$ in the case of
the resistive sheet and resistor-chains, respectively.
Upper limits obtained here indicate a promising potential of
using this material for rare-event-search experiments.
The $\alpha$-ray emission rate was also measured with an alpha-ray detector described in Ref.~\cite{bib:alpha} and the upper limit at 90$\%$ C.L was $\rm 2.4\times 10^{-2} \alpha /hour/cm^2$.
The $\alpha$-ray emission rate was less than that of the $\mu$-PIC
($\rm \left(3.57^{+0.35}_{-0.33}\right)\times10^{-1} \alpha /hour/cm^2$) described in Ref.~\cite{bib:alpha}. 
Dedicated measurements would be needed to confirm the requirement for each individual experiment.

\begin{table}[htbp]
\centering
\caption{\label{tab:HPGeresults} Measurement results with the HPGe. The values are shown in a unit of mBq/kg. Radioactivities of a typical resistor (KTR10EZPF, taken from Ref.~\cite{bib:HPGe} and re-normalized) are shown for reference. ${}^{226}$Ra and ${}^{228}$Ra are the isotopes in the ${}^{238}$U chain and ${}^{232}$Th chain, respectively.}
\smallskip
\begin{tabular}{|l|l|l|l|l|}
\hline
&
${}^{226}$Ra (${}^{238}$U chain)&
${}^{228}$Ra (${}^{232}$Th chain)&
${}^{40}$K&
${}^{60}$Co\\ \hline 
Achilles-Vynilas&
$< 18.4$&
$< 7.77$&
$< 112$&
$<2.54$\\ \hline 
Resistors KTR10EZPF &
$(4.1\pm0.5)\times10^2$ &
$(4.3\pm0.5)\times10^2$ &
$(4.2\pm0.6)\times10^3$ &
$< 25$\\ \hline
\end{tabular}
\end{table}

\begin{figure}[!h]
\centering 
\includegraphics[width=.6\textwidth]{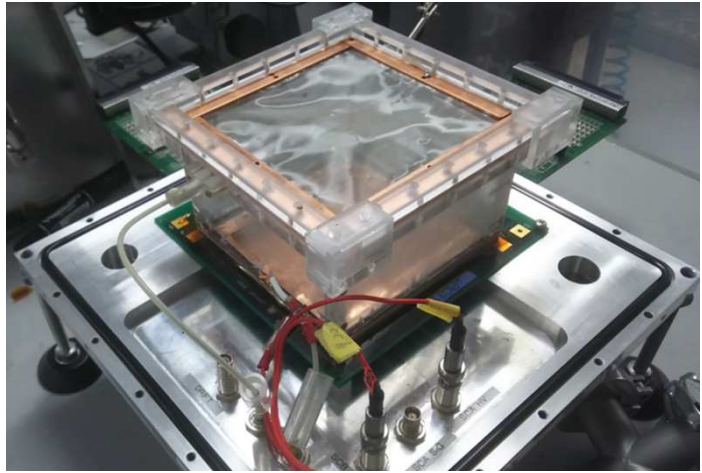}
\caption{\label{fig:SR_photo}Photo of the {\SR}. Sheet resistors are attached to the inner plane of the four walls made of acrylic.}
\end{figure}

\begin{figure}[!h]
\centering 
\includegraphics[width=1.\textwidth]{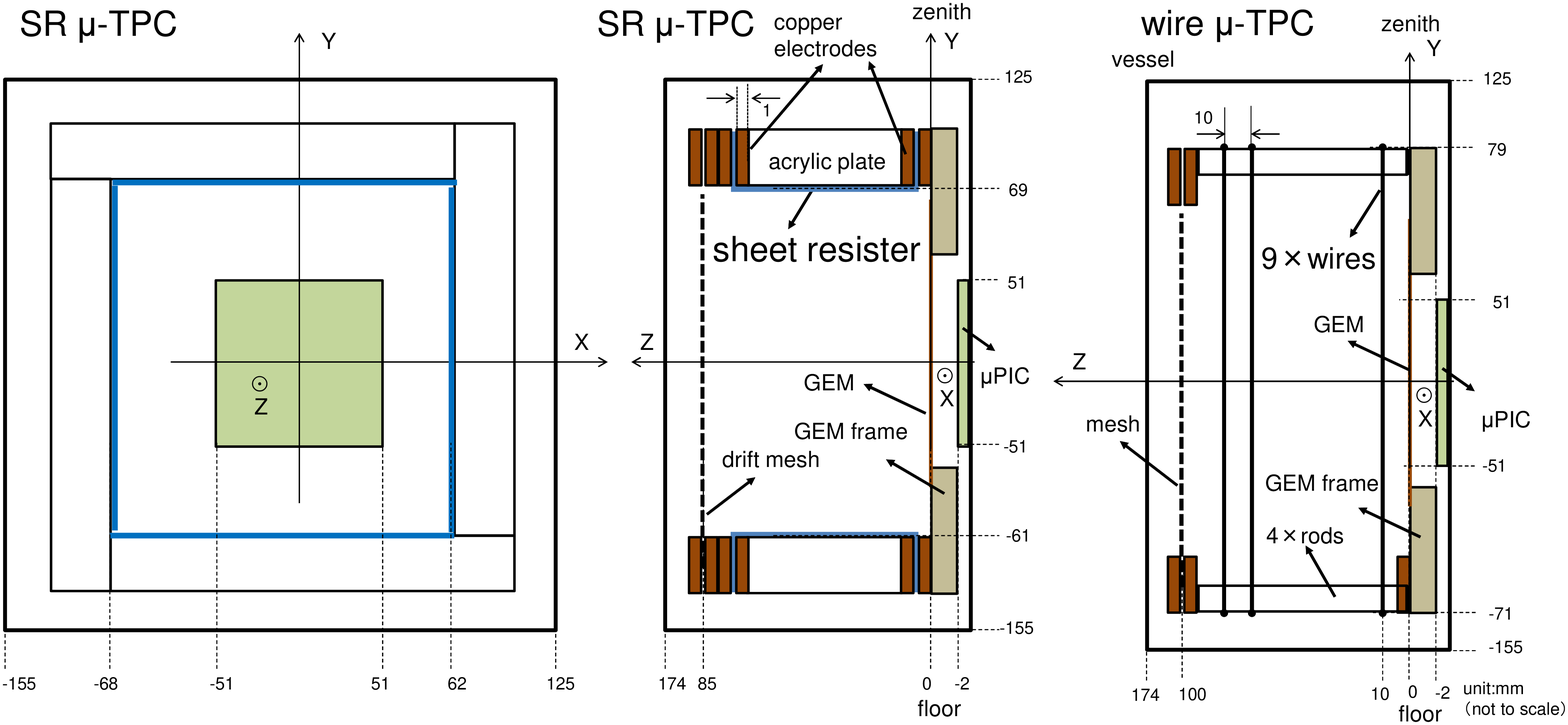}
\caption{\label{fig:TPC_schematic}Schematic drawings of the {\SR} and the {\wire}. Side views of the {\SR} and side view of {\wire} are shown at left (from drift-top), center (from drift-side), and right (from drift-side), respectively. Please note that these drawings are not to scale.} 
\end{figure}

A sheet-resistor field cage was made with Achilles-Vynilas.
A photo and schematic drawings of the {\SR} are shown in Fig.~\ref{fig:SR_photo} and Fig.~\ref{fig:TPC_schematic}, respectively.
The field cage was built by combining 
four acrylic plates as ''walls'' and a steel mesh as a ''drift top''.
The size of each acrylic plate was 80$\times$140$\times$10~$\rm mm^3$.
The resistive sheets were attached to the inner area of the acrylic plate by thermally pressing at a temperature at which the acrylic became soft but the
properties of the resistive sheet were not affected. 
The drift-top and drift-bottom parts of the resistive sheet were sandwiched by copper plates with a thickness of 1~mm. These electrodes were screwed into each other so that the ohmic contact to the resistive sheet was secured.
The inner size of the field cage was 130$\times$130~$\rm mm^2$ and the drift length was 85~mm.
The field cage was then coupled with the GEM + {\uPIC} gas detector system.
The GEM (made by Scienergy) was made of liquid crystal polymer with a thickness of 100~$\mu$m, a hole-diameter of 70~$\mu$m, and a hole-pitch of 140~$\mu$m.
The GEM was used as the first-stage gas amplifier.
The $\mu$-PIC was a two-dimensional readout device which had
orthogonally-placed strips with a pitch of 400~$\mu$m.
The charge from each strip was digitized by an amplifier-shaper-discriminator (ASD) chip\cite{bib:ASD} and synchronized with a field-programable-gate-array (FPGA)-based electronics board with a system frequency of 100MHz.
The start time and duration of the discriminated signal from each strip were recorded. Here the time reference was provided by an external trigger.
The {\uPIC} system thus recorded the two-dimensional images on two planes.
Details of the gas-detector system, with the difference in the detector size, can be found in Ref. \cite{bib:NEWAGE}.
The gas-amplification area of the GEM and the detection area of the {\uPIC} were
$\rm 100\times100$~$\rm mm^2$ and $\rm 102\times102$~$\rm mm^2$, respectively.
The detection volume of the {\SR} was thus $\rm 100\times100\times85$~$\rm mm^3$.
The origin of the coordination system was set at the center of the GEM so that the
XY plane corresponded the readout plane of the {\uPIC} and the Z axis
corresponded to the drift direction.
An existing field cage made of wire electrodes, {\wire}, was used for performance comparison. A schematic drawing of the {\wire} is shown in the right panel of Fig.~\ref{fig:TPC_schematic}. The inner size of the field cage was 150$\times$150~$\rm mm^2$ and the drift length was 100~mm. 
Due to the design of the $\mu$-PIC board, the centers of the field cage and the detection area were not aligned perfectly; there were offsets of 3~mm and 4~mm in the X and Y directions, respectively. The largest distance between the sheet-resistor field cage and the detection area was 18~mm (=69-51~mm, shown in the +Y area), as seen in the center drawing of the Fig.~\ref{fig:TPC_schematic}, while that of wire field cage was 29~mm.
The tracking-performances of the {\SR} and {{\wire}} in the regions more than 20~mm and 30~mm apart from the field cage were studied with these detectors.

\section{Performance test}
\label{sec:performance}

Performance tests of the {\SR} and {\wire} were conducted using cosmic-ray muons.
The chamber was filled with a gas mixture of argon (0.88~bar) and ethane (0.12~bar)
and operated with a total gas gain of 4.0$\times10^4$.
A drift field of 0.2~kV/cm was formed.
The XZ plane was set horizontally and the Y axis was aligned vertically as shown in Fig.~\ref{fig:TPC_schematic}. 
A coincidence of two plastic scintillators were used for the trigger.
Both scintillators were placed above the TPC chamber for the  {\SR} measurement (Y=130~mm and 170~mm) while one was set below the TPC for the {\wire} measurement(Y=-185~mm and 170~mm)\footnote{Measurements with the same trigger condition were performed but the gas conditions were found to differ so these two data-sets were used for the analysis.}.
The measurement periods after the gas filling were 0.2-0.32 days for the {\SR} and 0.1-0.52 days for the {{\wire}}.

Track events for the analysis were selected from the total event samples.
Low energy events below 5~keV 
were rejected in order to exclude noise events. The energy spectrum of the muon events peaked at 30~keV. Tracks with length between 7~cm and 15~cm were selected in order to exclude potential systematic error due to the trigger-condition difference. After these event selections, the event rates were $\rm (2.6\pm 0.2)\times 10^{-2} [events/s]$ and $\rm (2.3\pm 0.1)\times 10^{-2} [events/s]$ for the {\SR} and {\wire}, respectively. These rates were consistent within 2 $\sigma$ and the muon tracks with same properties were selected for the analysis.



 

\begin{figure}[htbp]
\centering 
\includegraphics[width=.6\textwidth]{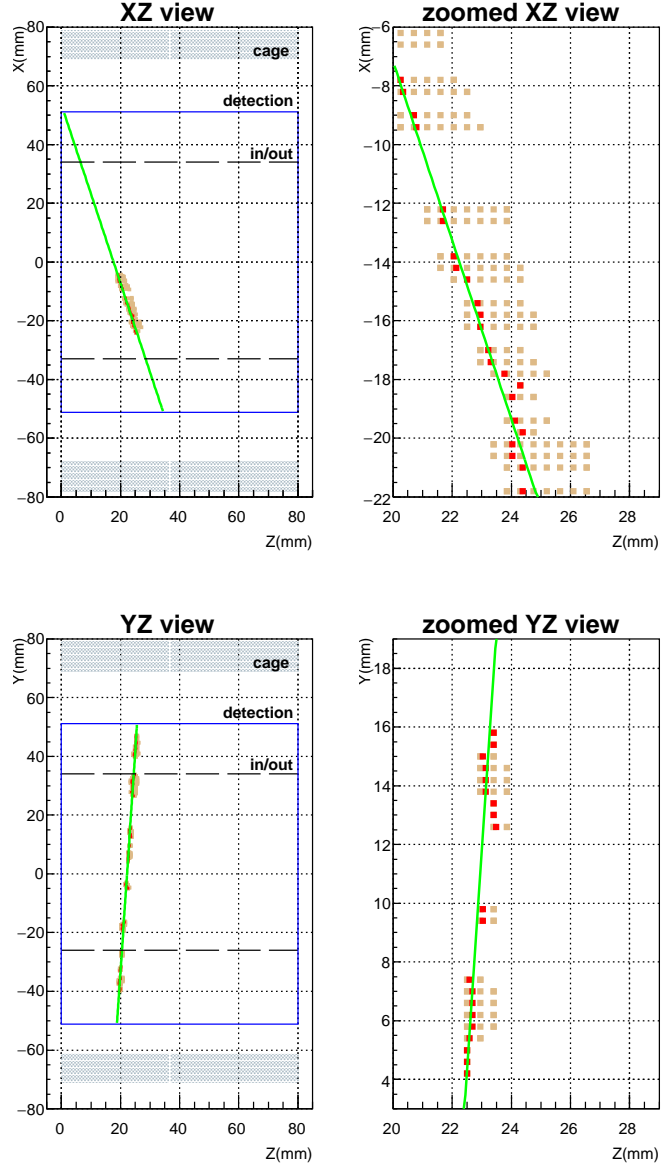}
\caption{\label{fig:eventdisplay}A typical muon event taken by the {\SR}. The upper and lower figures show the tracks in the XZ(top-view) and YZ(side-view) planes, respectively. The right figures show zoomed images of the same event. The pale-red marks are the discriminated signals from each strip and the red marks show the hits.  The green lines show the best-fit lines. The gray parts indicate the TPC cages. The solid and dashed lines represent the detection volume and boundary of the inner and outer regions, respectively.}
\end{figure}

A typical muon event detected with the {\uPIC} system is shown in Fig.~\ref{fig:eventdisplay}.
The discriminated signal of each strip is shown
by a pale-red mark in Fig.~\ref{fig:eventdisplay}.
The representative hit-timing (T) for each strip was determined at 20$\%$ of the signal duration from the beginning of the signal and it is shown with a red mark\footnote{This method was modified from the one introduced in Ref. \cite{ref:KomuraM}}. 
T was used to calculate Z based on the time from the trigger and the drift velocity (4.5~cm/$\mu$s, measured by the full-drift length hits).
Hereafter, this set of (X, Z) or (Y, Z) is referred to as a ''hit''.
The hits were fitted with straight lines independently in the XZ and YZ planes.
Best-fit results are shown with the green lines in Fig.~\ref{fig:eventdisplay}.
Each hit has measured two-dimensional coordinates (XZ or YZ) and the position of the third coordinate (Y or X) was calculated by the Z value and the line on the other plane so that each hit had a three-dimensional position.
The (X, Y) position was used to categorize each hit into the ''inner'' or ''outer'' region in order to evaluate the effect of the distance from the field cage. Here the inner region was defined as the area that is at least 3.5~cm apart from the field cage for both of the {\SR} and the {\wire}. This boundary was determined so that similar numbers of hits were expected in both outer and inner regions in total.
The inner and outer boundaries of the {\SR} are shown in Fig.~\ref{fig:eventdisplay} with dashed lines.
  
\begin{figure}[htbp]
\centering 
\includegraphics[width=1.\textwidth]{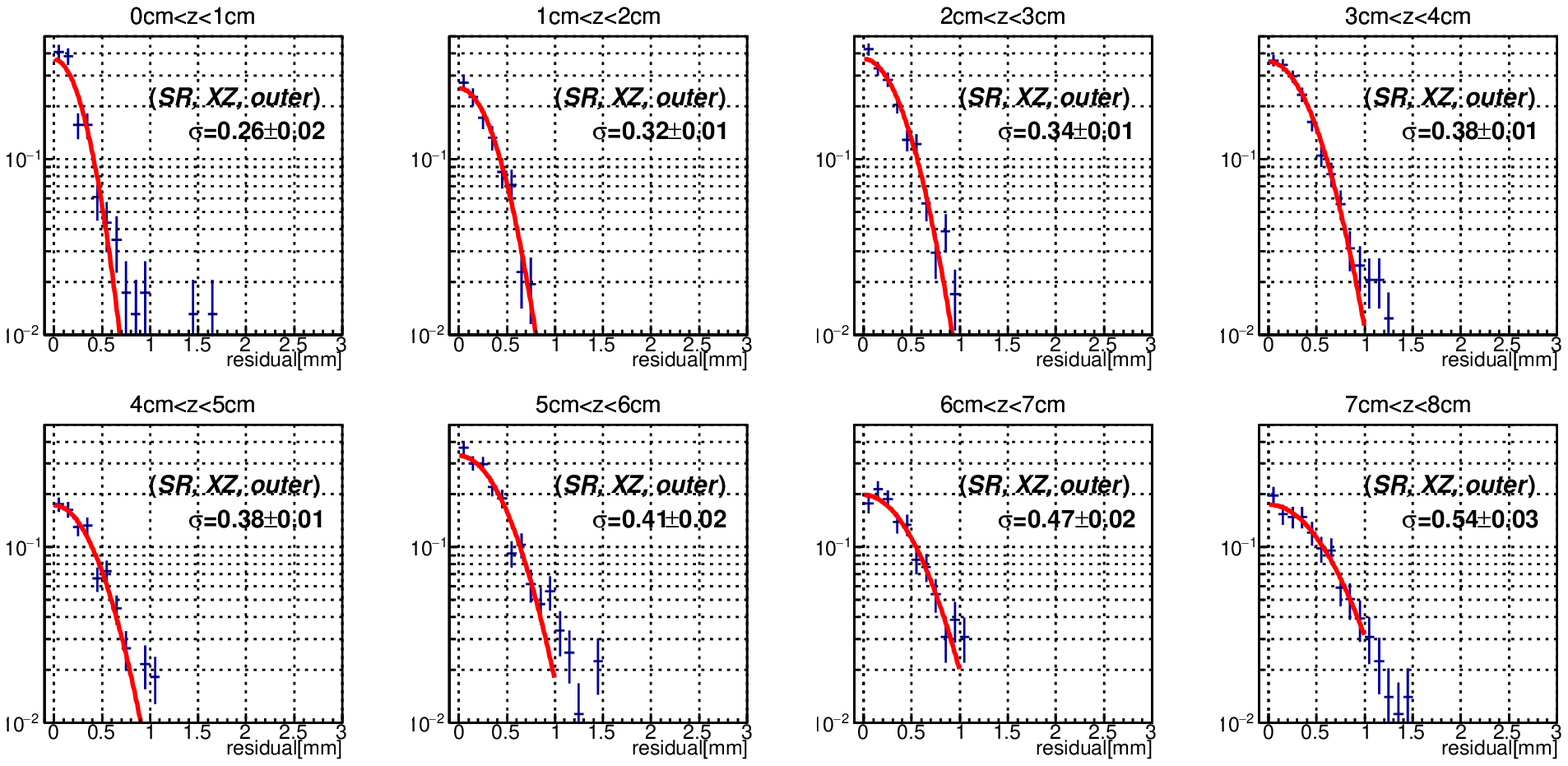}
\includegraphics[width=1.\textwidth]{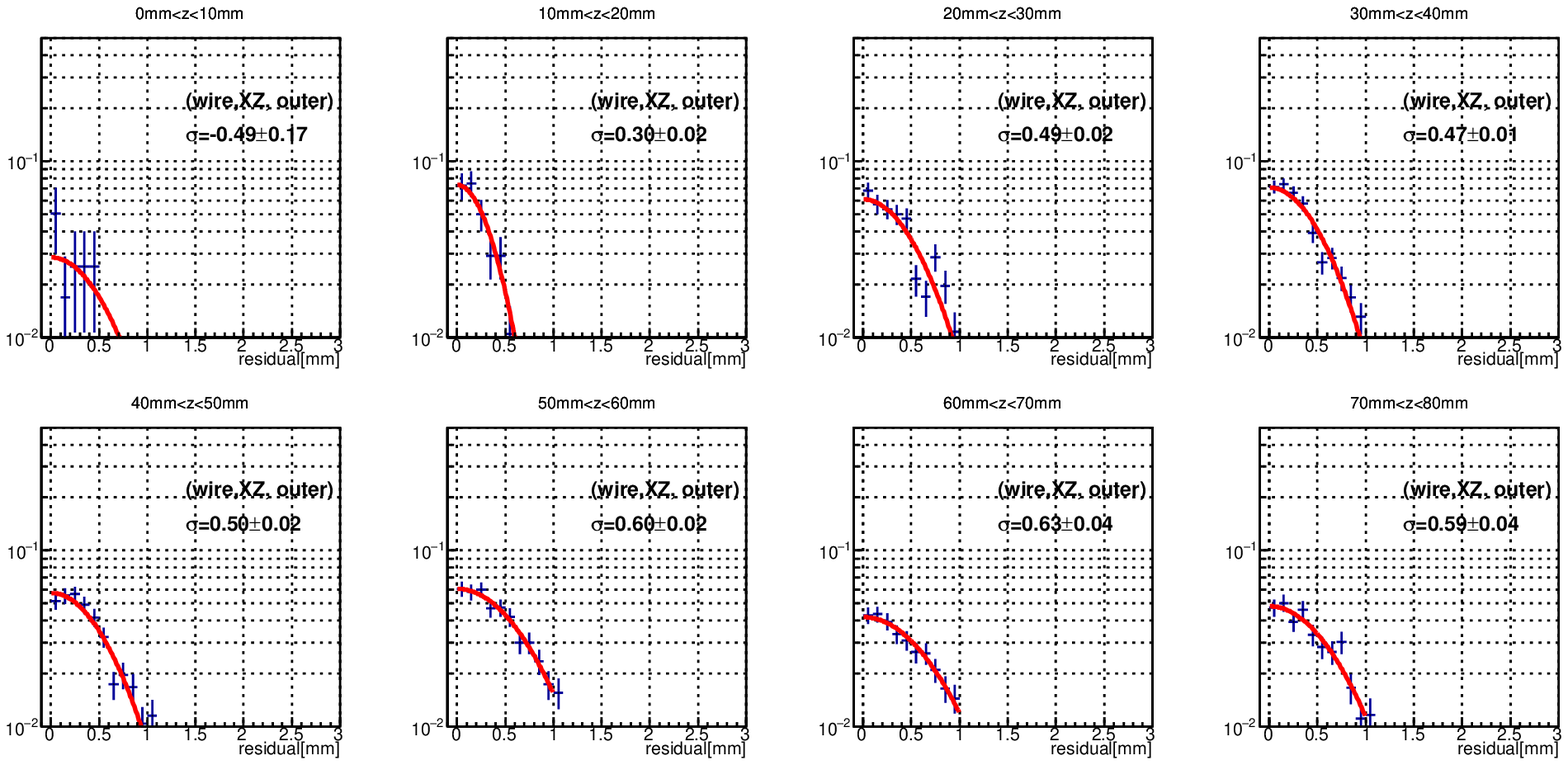}
\caption{\label{fig:residual_distributions}   The residual distributions of the ($SR, XZ, outer$) (upper) and ($wire, XZ, outer$) (lower) data. }
\end{figure}

Residuals were calculated as the distance of a hit and the best-fit line in each plane. According to the combinations of ($SR$ or $wire$) $\times$ ($XZ$ or $YZ$) $\times$ ($inner$ or $outer$), the data were independently analyzed and compared.
The combination is labeled as ($SR, XZ, inner$) for hits in the inner region analyzed on the XZ plane taken with the {\SR} and for other combinations in the same manner.
Typical residual distributions of the ($SR, XZ, outer$) and ($wire, XZ, outer$) data are shown in Fig.~\ref{fig:residual_distributions}. Here the events are classified according to Z or the drift distance.
Each residual distribution was fitted with a Gaussian function whose mean was fixed at 0. The height and the $\sigma$ were treated as free parameters. The results of the fitting are also shown in Fig.~\ref{fig:residual_distributions} with red lines.
The $\sigma$ of each distribution can be represented as Eq.~(\ref{eq:sigma_definition}).
\begin{equation}
  \label{eq:sigma_definition}
  \{\sigma_{i,j,k}(Z)\}^2=\{\sigma_{{\rm dd},(i,j,k)}\}^2+\{\sigma_{{\rm diff},(i,j,k)}(Z)\}^2
\end{equation}
Here, $i$= $SR$ or $wire$ is the field cage type, $j$=$XZ$ or $YZ$ is the detection plane, and  $k=$ $inner$ or $outer$ is the region of interest.
$\sigma_{\rm dd}$ comprises the detector-intrinsic and drift-field dependent position resolutions and is Z independent. $\sigma_{\rm diff}$ is a diffusion-related term which has a Z-dependence as Eq.~(\ref{eq:sigma_definition_sigma_diff}).
\begin{equation}
  \label{eq:sigma_definition_sigma_diff}
  \sigma_{{\rm diff},(i,j,k)}(Z)=d'_{j}\sqrt{Z},
\end{equation}
where $d'_{j}$ is an effective-diffusion in mm at $Z$=1~cm. Since the effective number of primary electrons for the determination of one hit points is practically more than one, measured $d'_j$ parameters are better than the diffusion values defined by the expected position for a single electron.
It should be noted that $\sigma_{\rm diff}$ basically depends upon the electron diffusion in the gas and is therefore independent of the field-cage type $i$ and position $k$. The dependence on $j$  still remains, owing to the track directions. For the YZ plane case, transverse diffusion, which is in the Y-axis direction, may be observed smaller than the original size 
because the most of the track directions are also in the Y-axis directions. 
This effect is also quantitatively treated in the next section.

\begin{figure}[htbp]
\centering 
\includegraphics[width=1.\textwidth]{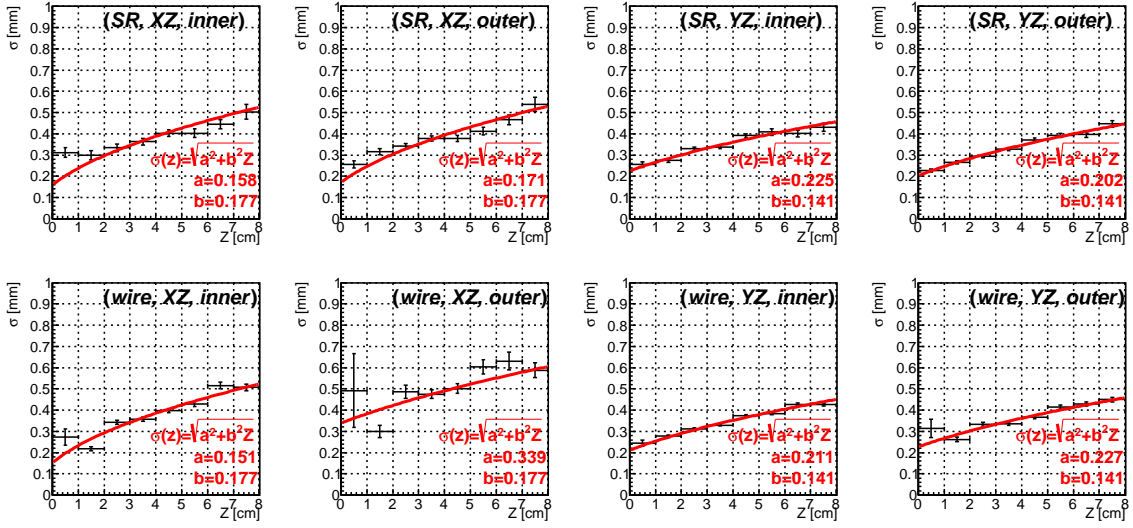}
\qquad
\caption{\label{fig:residual_comparison} $\sigma$ dependence on Z for eight data-sets.}
\end{figure}
\begin{figure}[htbp]
\centering 
\includegraphics[width=1.\textwidth]{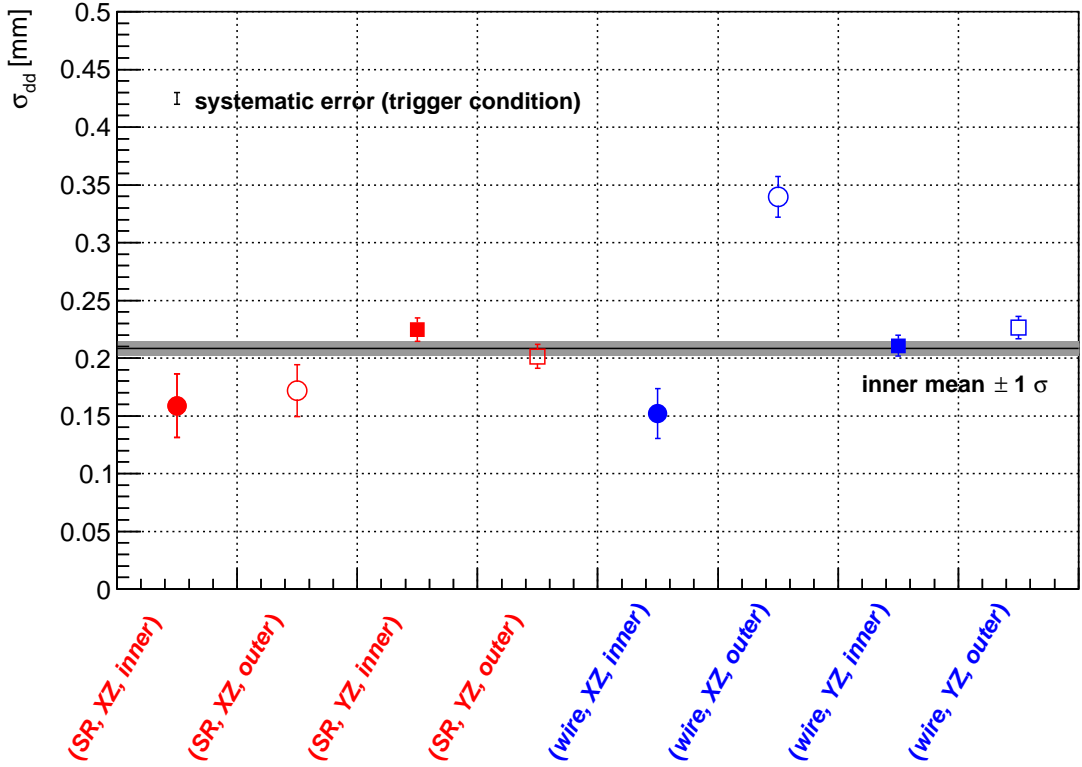}
\qquad
\caption{\label{fig:finalresults} Obtained $\sigma_{dd}$. Red and blue markers show {\SR} and {\wire} results, respectively. Circles and squares show XZ and YZ results, respectively. Filled and open markers show the inner and outer results, respectively.}
\end{figure}

$\sigma$ parameters derived from the fitting of the distributions of the residuals.
They were plotted as a functions of Z and are shown in Fig.~\ref{fig:residual_comparison}. There, eight results corresponding to eight data-sets are all shown. All of the data (64 bins) were simultaneously-fitted with Eq.~(\ref{eq:sigma_definition}). Here, 10 (8 for $\sigma_{{\rm dd},(i,j,k)}$ + 2 for $d_{j}$) free parameters were used. The results are shown with red lines and characters in Fig.~\ref{fig:residual_comparison} and also in Table~\ref{tab:fitresults}. The $\chi^2$/NDF of the fitting was 60.2/54. Obtained $\sigma_{{\rm dd}}$ values are shown in Fig.~\ref{fig:finalresults}. There, the eight conditions are labeled below the data.
The size of the systematic error due to the difference of the trigger condition is indicated with a solid line on the left-top, which will be discussed in the following section.

\begin{table}[htbp]
\centering
\caption{\label{tab:fitresults} Fit results for the $\sigma_{{\rm dd},(i,j,k)}$ and $d'_{j}$. The units are mm for $\sigma_{{\rm dd},(i,j,k)}$ and $\rm mm/\sqrt{cm}$ for $d'_{j}$.}
\smallskip
\begin{tabular}{l|l}
\hline
parameter& best-fit value\\ \hline
  $\sigma_{{\rm dd},(SR,XZ,inner)}$& 0.158    $\pm$    0.043\\
  $\sigma_{{\rm dd},(SR,XZ,outer)}$&   0.171    $\pm$     0.031  \\
  $\sigma_{{\rm dd},(SR,YZ,inner)}$&  0.225    $\pm$     0.013\\
  $\sigma_{{\rm dd},(SR,YZ,outer)}$&  0.202    $\pm$     0.013  \\
  $\sigma_{{\rm dd},(wire,XZ,inner)}$&  0.151     $\pm$    0.035\\
  $\sigma_{{\rm dd},(wire,XZ,outer)}$&   0.339    $\pm$     0.021 \\
  $\sigma_{{\rm dd},(wire,YZ,inner)}$&   0.211     $\pm$    0.013\\
  $\sigma_{{\rm dd},(wire,YZ,outer)}$&    0.227     $\pm$    0.013 \\ \hline
$d'_{XZ}$ &0.177$\pm$ 0.007\\
$d'_{YZ}$ & 0.141$\pm$ 0.003\\ \hline

\end{tabular}
\end{table}

The primary result of this work was that the {\SR} actually worked. This was demonstrated by Fig.~\ref{fig:eventdisplay} and the $SR$ data in Fig.~\ref{fig:finalresults} being comparable to the ($wire, inner$) data. 
Thus, it can be said that 
no significant differences between ($SR, XZ, inner$) and ($SR, XZ, outer$),  ($SR, YZ, inner$) and ($SR, YZ, outer$) were seen. It meant that a good performance was kept 
at the outer region (20~mm from the field cage) of the {\SR}.
The mean of the inner-data and their 1 $\sigma$ band are shown by the solid line and gray band for reference. No significant differences were observed between the ($SR, outer$) data and the $inner$ mean. 
This result can be highlighted when the ($wire, XZ, outer$) result was compared with other data.
The result was significantly larger than the others,  
indicating that the electric field of the {\wire} was deteriorated due to the ground-voltage of the vessel seen through the wires. The ($wire, XZ, inner$), on the  other hand, showed a result consistent with the other data, confirming that there was no systematic problem for the ($wire, XZ$) setup. 
The comparison of ($SR, XZ, outer$) and ($wire, XZ, outer$) illustrates that the sheet-resistor field cage was useful for shielding the potential of the vessel.
The difference between ($wire, XZ, outer$) and ($wire, YZ, outer$) can be due to the
track direction and the diffusion, which will be discussed in Section~\ref{sec:discussion}.

\section{Discussion}
\label{sec:discussion}
Our measurements indicate promising results for the performance of the {\SR}.
Although the comparison-based results discussed in Section~\ref{sec:performance}
were robust and reliable,
some discussion would be important to confirm the consistency and correctness of the all measurements.

First, as summarized in Table~\ref{tab:fitresults}, the measured effective-diffusion on the XZ plane ($d'_{\rm XZ}$=0.177$\pm$ 0.007) was larger than that on the YZ plane ($d'_{\rm YZ}$=0.141$\pm$ 0.003).
This result can be explained by the following discussion.
The transverse and longitudinal diffusions, $d_{\rm t}$ and $d_{\rm l}$,
calculated with MAGBOLTZ~11.6 simulation\cite{bib:MAGBOLTZ},
were $0.48~{\rm mm}/\sqrt{\rm cm}$ and $0.29~{\rm mm}/\sqrt{\rm cm}$, respectively.
Here the transverse diffusion can be observed as the diffusions along the X and Y axises and the longitudinal one can be seen as the diffusion along the Z axis. It should be noted that when the diffusions are measured using long tracks the diffusions along the track directions are not observed but only the diffusions perpendicular to the track can be observed.
This effect can be written as Eqs.~(\ref{eq:phitheta})-(\ref{eq:phitheta2}). 

\begin{eqnarray}  \label{eq:phitheta}
(d_{\rm XZ}(\phi))^2&=&(d_{\rm t}\sin\phi)^2+(d_{\rm l}\cos\phi)^2\\ \label{eq:phitheta2}
(d_{\rm YZ}(\theta))^2&=&(d_{\rm t}\sin\theta)^2+(d_{\rm l}\cos\theta)^2,
\end{eqnarray}

where $d_{\rm XZ}(\phi)$ and $d_{\rm XZ}(\theta)$ are the observable diffusions on the XZ and YZ planes, respectively. $\phi$ and $\theta$ are the angle between the track and the X and Y axises, respectively. The detector was set so that the $\phi$ was the azimuth angle and the $\theta$ was the zenith angle.
The weighted means of  Eq.~(\ref{eq:phitheta}) and Eq.~(\ref{eq:phitheta2}), $d_{\rm XZ}$ and $d_{\rm YZ}$, 
were calculated by simulation taking account of the detector geometry and the zenith angle dependence of the cosmic muon. Obtained values were $d_{\rm XZ}=0.39~{\rm mm}/\sqrt{\rm cm}$ and  $d_{\rm YZ}=0.32~{\rm mm}/\sqrt{\rm cm}$.
These observable diffusions were related to 
effective diffusions as Eqs.~(\ref{eq:effectiveandobservable})-(\ref{eq:effectiveandobservable2}). 

\begin{eqnarray}  \label{eq:effectiveandobservable} 
d'_{\rm XZ}&=&d_{\rm XZ}/\sqrt{N_{\rm XZ}}\\ \label{eq:effectiveandobservable2} 
d'_{\rm YZ}&=&d_{\rm YZ}/\sqrt{N_{\rm YZ}},
\end{eqnarray}
where $N_{\rm XZ}$ and $N_{\rm YZ}$ are the effective number of electrons for the determination of each hit point in the ZX and YZ plane, respectively.
The result can be explained with $N_{\rm XZ}{\sim}N_{\rm YZ}{\sim}5$. The difference between $\sigma_{{\rm dd},(wire,XZ,outer)}$ and $\sigma_{{\rm dd},(wire,YZ,outer)}$ can be understood in the same way; smaller values are obtained for the position displacement along the track direction (Y axis) than that along the X axis. In this case, measured values can be explained with $\sigma_{{\rm drift},(wire,Z,outer)}=0.1~{\rm mm}$ and $\sigma_{{\rm drift},(wire,XY,outer)}=0.5~{\rm mm}$, where $\sigma_{{\rm drift},(wire,Z,outer)}$ and $\sigma_{{\rm drift},(wire,XY,outer)}$ are standard deviations of the  position displacements by the drift field distortion in the Z direction and XY plane, respectively. Precise calculation on the electric field in the future work would provide a more quantitative explanation.

Here it was shown the track direction affected the $\sigma$ results. Therefore the trigger condition difference between the two measurements could cause systematic errors although the track-length cut provided a good event selection in terms of the event rate. A parameter, $\Delta X$, the difference between the maximum and minimum $X$ positions in the hits in a track was used to evaluate the systematic error because this parameter represents the track direction. The mean values of $\Delta X$ of the selected events in the two measurements, $\overline{\Delta X(SR)}$ and $\overline{\Delta X(wire)}$ were 1.6~cm and 1.4~cm, respectively. When the events with $\Delta X > 3~{\rm cm}$ were excluded in the $SR$ data, $\overline{\Delta X(SR)}$ was 1.3~cm which was smaller than that of $\overline{\Delta X(wire)}$. The difference of the $\sigma_{{\rm dd},(SR)}$ between the original result and the one with this additional cut was treated as the systematic error and is indicated in Fig.~\ref{fig:finalresults}. The systematic error due to the trigger condition difference was confirmed to be smaller than the statistic errors and was found not to affect the conclusions of this work.

Second, it is interesting to know if a drift-field dependent term can be separated from the detector-intrinsic since the results shown in Table \ref{tab:fitresults} are all convoluted values. After some trial, this data set was found to be not enough to give any conclusive results on the drift-field dependent term.

Finally, let us discuss the remaining studies that need to be done on {\SR} before a practical use in a large-scale TPC. The tracking-performance in the area 20~mm or more away from the field cage was confirmed. One of the potential advantages of {\SR} is that there would be less
electric-field deterioration near the wall than with a ring-type or tape-type TPCs. A dedicated field cage with the same or smaller area compared to the detection area would help for a further study on this potential. 
Long-term-stability is also of particular interest, since rare-event-search experiments usually require stable measurements on the order of years.
Another interesting study is the use in cryogenic systems for the liquid-noble-gas TPCs. Unfortunately, this particular product (Achilles-Vynilas) showed a semiconductor-like temperature dependence ($\rm \propto exp(-1.5\times10^4 T)$) at the room temperature.
It is thus expected that the resistivity would be too high at a temperature below 100~K, which suggests it is necessary to find another material. The resistive material for MPGD would be a solution for the cryogenic version of {\SR}\footnote{There was a report after the submission of this work on a ''Resistive Shell Liquid Argon TPC''\cite{bib:RS-Ar}}.

\section{Conclusion}
\label{sec:conclusions}
A new-concept TPC using a commercial resistive sheet, {\SR},
was developed and its performance was measured.
With a sheet resistor, {\SR} has the potential to make a more uniform electric field than the conventional TPCs with resistor-chains. 
Detector assembly was easier than that of conventional TPCs.
The material used in this study, Achilles-Vynilas, was found to be thin, transparent, and low-radioactive.
The tracking-performance measurement with cosmic muons showed 
very promising results, indicating this type of field cage actually worked and showed 
a good tracking-performance even in the volume close to (20~mm) the field cage.
This type of TPC field cage offers an alternative for the widely used conventional field cages.

\section*{Acknowledgment}

We gratefully acknowledge
the cooperation of Kamioka Mining and Smelting Company.
We thank the XMASS collaboration for their help on the low-background
measurement technologies.
This work was supported by the Japanese Ministry of Education,
Culture, Sports, Science and Technology, Grant-in-Aid
for Scientific Research, ICRR Joint-Usage, JSPS KAKENHI
Grant Number, 16H02189 26104004 26104005 26104009, and
JSPS Bilateral Collaborations (Joint Research Projects and Seminars) program.


%

\end{document}